\documentclass[preprint,tightenlines,eqsecnum,floats,aps,amsmath,amssymb,nofootinbib,prd,showpacs]{revtex4}

\usepackage{amssymb}
\usepackage{stmaryrd}
\usepackage{amsmath}
\usepackage{amsfonts}
\usepackage{mathrsfs}
\usepackage{CJK}
\usepackage{amsmath,amssymb,amsfonts}
\usepackage{graphicx}
\usepackage{subfigure}

\def\be{\begin{equation}}
\def\ee{\end{equation}}
\def\ba{\begin{eqnarray}}
\def\ea{\end{eqnarray}}
\def\nn{\nonumber}

\begin{document}

\title{Kerr-de Sitter and Kerr-anti-de Sitter black holes as accelerators for spinning particles }

\author{Songming Zhang, Yunlong Liu}
\affiliation{Department of Physics, South China University of Technology, Guangzhou 510641, China}

\author{ Xiangdong Zhang\footnote{Corresponding author. scxdzhang@scut.edu.cn}}
\affiliation{Department of Physics, South China University of Technology, Guangzhou 510641, China}

\date{\today}


\begin{abstract}
It is shown that the Kerr black hole could act as accelerators for spinning particles. In principle, it could obtain arbitrarily high energy for extremal Kerr black hole. In this paper, we extend the previous research to the Kerr-(anti-)de Sitter background and find that the cosmological constant plays
an important role in the result. For the case of Kerr-anti-de Sitter
black holes, like the Kerr background, only the extremal Kerr-anti-de Sitter
black holes can have a divergent center-of-mass energy of collision.
While, for the case of Kerr-de Sitter black holes, the collision of two spinning particles can take place on the outer
horizon of the black holes and the center-of-mass energy of collision can blow
up if one of the collision particle takes the critical angular momentum. Hence, non-extremal
Kerr-de Sitter black holes could also act as  accelerators with arbitrarily
high center-of-mass energy for spinning particles.

\end{abstract}

\

\maketitle

\section{Introduction}
In 2009, Ba\~nados, Silk and West(BSW) firstly showed that rotating black hole can act as particle accelerators\cite{111102}. They showed that for critical Kerr black hole, the collision center-of-mass energy can be arbitrarily high if two test particles rest at infinity collide near the event horizon\cite{111102}. Along this line, a lot of progress have been made in the decade to further study the issue of BSW mechanism\cite{103005,675,044013,023004,024016,025008,280,105014,084025,124017,064035,58.064005,124030,507,1009,225006,406,084041,083004}. So far, most authors focus on  point particles which trajectory is a geodesic. However a real particle is an extended body with
self-interaction. It has been shown \cite{209248,314499,6406,237} that the trajectory
of a spinning test particle is no longer a geodesic. And it orbits of spinning particles
around black holes background have governed by the
Mathisson-Papapetrou-Dixon(MPD) equations \cite{064035,58.064005,124030}.
By employing MPD equations, in\cite{084025}, the authors show that for an extremal Kerr black hole, the collision energy could be divergent with some additional critical condition be satisfied.
However, as show in \cite{507}, the spin of astrophysical black holes should be less than $0.998M$ ($M$ is the mass of the black hole) which means there is no extremal Kerr black hole existed in nature.

In the past years, people have found many evidences
coming from the cosmological observation which shows our present universe is in a state of accelerated expansion. Although many plausible models
are constructed to explain the existence of such accelerated expansion, most observations favor the cold dark matter model with a cosmological constant ($\Lambda$-CDM model)\cite{1009}. At least phenomenologically the Einstein equations should be modified with a cosmological constant $\Lambda$ at cosmological scale. With a cosmological constant, the Kerr black hole should be
generalized to the Kerr-de-Sitter(Kerr-dS) background.

On the other hand, the AdS/CFT correspondence becomes a fruitful field, and had many interesting results has been made in the past decades\cite{TASI,Kerrcft,59.064005}. In\cite{Hebecker13}, the authors investigate the issues of
CFT dual to collision particles on a given black hole background. Therefore, it is also interesting to extend Kerr black hole to Kerr-anti-de Sitter background and study the process of collisions of spinning particles. Moreover, in\cite{225006}, the authors showed that unlike the Kerr case, for the Kerr-dS black hole, even the non-extremal Kerr-dS background can serve
as accelerator for point particles without spin, and its corresponding collision energy can be divergent when some critical conditions satisfied.

With all these strong motivations in hand, in this paper, we study the possibility of Kerr-de Sitter and Kerr-anti-de Sitter black holes as accelerator for spinning particles. By using the MPD equations, we investigate the BSW process of the Kerr-(anti-)de Sitter black hole. Because the Kerr-(anti-)de Sitter black hole
have a more complicated horizon structure than Kerr black holes, it is very hopeful that some novel features will be emergent.

The paper is organized as follows. In section \ref{Equation}, we introduce the equations of motion for spinning particles. In section \ref{spnning}, the four momentum of a spinning particle are solved in Kerr-(anti)-dS background. In section \ref{CM energy}, we obtained the collision center-of-mass energy of two spinning particles. And then we discuss the Kerr-dS and Kerr-AdS background separately in section \ref{Kerr-dS} and section \ref{Kerr-AdS}. The summary and conclusion are given in section \ref{conclusion}.

Through out the paper, we adopt convention that gravitational constant $G$ and the speed of light $c$ equal to unity.

\section{Equations of motion for spinning particles}\label{Equation}

The trajectory of a spin particle in the curved spacetime is described by the Mathission-Papapetrou-Dixon(MPD) equations\cite{406,084025}

\begin{eqnarray}
&\frac{DP^{a}}{D\tau }=-\frac{1}{2}R_{bcd}^{a}\upsilon ^{b}S^{cd},\nn
\end{eqnarray}
\begin{eqnarray}
&\frac{DS^{ab}}{D\tau }=P^{a}\upsilon ^{b}-P^{b}\upsilon ^{a}. \label{2.1}
\end{eqnarray}
where
\begin{equation}
\upsilon ^{a}=(\frac{\partial }{\partial \tau })^{a} \label{2.2}
\end{equation}
is the tangent vector of the center-of-mass world line,  $\frac{D}{D\tau }$ is the covariant derivative along $\upsilon ^{a}$, and $P^{a}$ is the canonical 4-momentum of the spinning particles satisfying the condition
\begin{equation}
m^{2}=-P^{a}P_{a}.
\end{equation}
Moreover $S^{ab}$ is the antisymmetric spin tensor which its square related to the spin and mass of the particle as follows \cite{58.064005}
\begin{equation}
\frac{1}{2}S^{ab}S_{ab}=S^{2}=m^2s^2,
\end{equation}
here $m$ and $s$ represent the mass and spin of the particles respectively.
In order to simplify the calculation in the main text and easy to gain the physical insight, people usually working in a specific frame which only 3-components of the spin tensor is nonzero\cite{105014}. This add to the spin supplementary condition\cite{105014,084025}
\begin{equation}
S^{ab}P_{b}=0
\end{equation}
or equivalently set $S^{0i}=0$. Again for latter convenience, we normalize the parameter $\tau$ in Eq. (\ref{2.2}) as,
\ba
u^av_a=-1
\ea
which means $\tau$ is not the proper time of the spin particle. The detailed calculation shows the relation between $u^{a}$ and $v^{a}$ can be writen as \cite{58.064005,084025}
\be
v^{a}-u^{a}=\frac{S^{ab}R_{bcde}u^{c}S^{de}}{2(m^{2}+\frac{1}{4}R_{bcde}S^{bc}S^{de})}. \label{2.7}
\ee
Furthermore, for the spacetime with a Killing vector field $\xi^{a}$, we can define the following conserved quantity for spinning particles,
\begin{equation}
Q_{\xi }=P^{a}\xi _{a}-\frac{1}{2}S^{ab}\triangledown _{b}\xi ^{a} \label{2.8}
\end{equation}
which is very helpful to find the trajectory of the spinning particle.

\section{spinning particles in Kerr-AdS and Kerr-dS background}\label{spnning}

In this section, we focus ourself on the Kerr-dS and Kerr-AdS case. The corresponding spacetime metric in the Boyer-Lindquist coordinates is\cite{084041}

\begin{equation}
ds^{2}=-\frac{\Delta _{r}}{\Sigma }(dt-\frac{asin^{2}\theta }{\Xi }d\varphi )^{2}+\frac{\Sigma }{\Delta _{r}}dr^{2}+\frac{\Sigma }{\Delta _{\theta}}d\theta ^{2}+\frac{\Delta _{\theta} sin^{2}\theta}{\Sigma }(adt-\frac{r^{2}+a^{2}}{\Xi }d\varphi )^{2} \label{3.1}
\end{equation}
where
\ba
\Delta _{r}&=&(r^{2}+a^{2})(1-\frac{\Lambda}{3}r^2)-2Mr, \label{3.2}\\
\Sigma &=&r^{2}+a^{2}cos^{2}\theta,\\
\Delta _{\theta }&=&1+\frac{\Lambda a^2}{3}cos^{2}\theta,\\
\Xi &=&1+\frac{\Lambda a^2}{3},
\ea
here parameters $M$ and $a$ correspond to the mass and angular momentum per unit rest mass of the black hole, $\Lambda$ is the positive(negative) cosmological constant.\\
The tetrad reads

\ba
e_{a}^{(0)}&=&\sqrt{\frac{\Delta_{r} }{\Sigma }}(dt_a-\frac{asin^2\theta }{\Xi }d\varphi _a),  \label{3.6}\\
e_{a}^{(1)}&=&\sqrt{\frac{\Sigma }{\Delta _r}}dr_a,\label{3.7} \\
e_{a}^{(2)}&=&\sqrt{\frac{\Sigma }{\Delta _\theta}}d\theta_a,\label{3.8} \\
e_{a}^{(3)}&=&\sqrt{\frac{\Delta _\theta}{\Sigma}}sin\theta(-adt_a+\frac{r^2+a^2}{\Xi }d\varphi_a). \label{3.9}
\ea

The stationary and rotational symmetry properties of the Kerr-dS(AdS) metric are characterized by two Killing vector fields:
the timelike Killing vector $\xi ^a=\left(\frac{\partial}{\partial t}\right)^a$ and the axial Killing vector $\phi^a=\left(\frac{\partial}{\partial \phi}\right)^a$.
For the convenient of the following calculation, we expanded these two Killing vectors in the tetrad formalism as
\ba
\xi_a&=&-\left(\sqrt{\frac{\Delta_{r} }{\Sigma }}e_{a}^{(0)}+\sqrt{\frac{\Delta_{\theta} }{\Sigma }}asin\theta e_{a}^{(3)}\right),\nn\\
\phi_a&=&\sqrt{\frac{\Delta_{r} }{\Sigma }}e_{a}^{(0)}+\sqrt{\frac{\Delta_{\theta} }{\Sigma }}(r^2+a^2)sin\theta e_{a}^{(3)}.
\ea
They are two conserved quantities correspond to these two Killing vectors, namely the energy of per unit mass of the particle $e=\frac{E}{m}$, and the $z$ component of total angular momentum per unit mass of the particle $j=\frac{J}{m}$. By applying Eq. (\ref{2.7}) and the tetrad formalism we have
\ba
e&=&-u^a\xi _a+\frac{1}{2m}S^{ab}\triangledown _b\xi _a,\nn\\
j&=&u^a\phi _a-\frac{1}{2m}S^{ab}\triangledown _b\phi _a. \label{3.11}
\ea
For simplicity, we only consider the situation where a spinning particle moves on the orbits in
the equatorial plane $(\theta =\frac{\pi }{2})$. Firstly, we introduce a special spin vector $s^{(a)}$ as\cite{58.064005,084025}
\begin{equation}
s^{(a)}=-\frac{1}{2m}\varepsilon _{~~~(b)(c)(d)}^{(a)}u^{(b)}S^{(c)(d)},
\end{equation}
or equivalently
\begin{equation}
S^{(a)(b)}=m\varepsilon ^{(a)(b)}_{~~~~~~(c)(d)}u^{(c)}s^{(d)},
\end{equation}
where $\varepsilon _{(a)(b)(c)(d)}$ is the completely anti-symmetric tensor with the component $\varepsilon _{(0)(1)(2)(3)}=1$. And the only non-vanishing component of $s^{(a)}$ to be\cite{58.064005,084025}
\begin{equation}
s^{(2)}=-s,
\end{equation}
where $s$ indicates not only the magnitude of spin but also includes the spin direction. The
particle spin is parallel to the block hole spin for $s>0$, while it is anti-parallel for $s<0$. Therefore the remaining non-vanishing tetrad components of the spin angular momentum are
\ba
S^{(0)(1)}&=&-msu^{(3)},\nn\\
S^{(0)(3)}&=&-msu^{(1)},\nn\\
S^{(1)(3)}&=&-msu^{(0)}. \label{3.15}
\ea
By calculating the tetrad components of Eq. (\ref{3.15}) and substitute it to Eq. (\ref{3.11}), we obtain the expression of the energy and
the angular momentum per unit mass $e$ and $j$ as
\ba
e&=&\frac{\sqrt{\Delta _r}}{r}u^{(0)}+\frac{a}{r}u^{(3)}+\frac{M-\frac{\Lambda }{3}r^3}{r^2}su^{(3)}, \label{3.16}\\
j&=&\frac{a\sqrt{\Delta _r}}{r}u^{(0)}+\frac{r^2+a^2}{r}u^{(3)}+\frac{a(M+r-\frac{\Lambda }{3}r^3)}{r^2\Xi }su^{(3)}+\frac{\sqrt{\Delta _r}}{r\Xi }su^{(0)} \label{3.17}
\ea
Solving the Eq. (\ref{3.16}) and Eq. (\ref{3.17}) gives
\ba
u^{(0)}&=&\frac{\tilde{K}}{\sqrt{\Delta _r}K}, \label{3.18}\\
u^{(3)}&=&\frac{\bar{K}}{K}. \label{3.19}
\ea
where
\ba
K&=&-3Ms(3s+a^3\Lambda )+r^3(9+3a^2\Lambda +3s^2\Lambda +a^3s\Lambda ^2), \label{3.20}\\
\tilde{K}&=&r(-j(3+a^2\Lambda )(3ar+3Ms-r^3s\Lambda ) \nn\\
&&+3e(3r^3+a^4r\Lambda +a^2r(3+r^2\Lambda )+as(3M+3r-r^3\Lambda ))), \label{3.21}\\
\bar{K}&=&3r^2(j(3+a^2\Lambda )-e(3a+3s+a^3\Lambda )). \label{3.22}
\ea

Note that we are working on the equatorial plane $(\theta =\frac{\pi }{2})$, so $u^{(2)}=0$. The normalization condition of $u^{(a)}u_{(a)}=-1$ gives us
\be
-(u^{(0)})^{2}+(u^{(1)})^2+(u^{(3)})^2=-1. \label{3.23}
\ee
By substituting Eq. (\ref{3.18}) and Eq. (\ref{3.19}) to Eq. (\ref{3.23}), we obtain
\be
(u^{(1)})^2=\frac{\tilde{K}^2-\Delta _r(\bar{K}^{2}+K^{2})}{\Delta _rK^2}. \label{3.24}
\ee

\section{CENTER-OF-MASS ENERGY}\label{CM energy}
Now we turn to the center-of-mass energy of the collision particles. For simplicity, we consider the two equal-mass particles with masses $m_{1}=m_{2}=m$. The collision center-of-mass energy of two spinning particles falling from infinity at rest is\cite{084025}
\ba
E_{cm}=\sqrt{2}m\sqrt{1-g_{ab}u_{(1)}^{a}u_{(2)}^{b}}, \label{4.1}
\ea
Substituting Eq. (\ref{3.18}), Eq. (\ref{3.19}) and Eq. (\ref{3.22}) to Eq. (\ref{4.1}), one can easily obtain
\ba
-g_{ab}u_{(1)}^{a}u_{(2)}^{b}&=&u_{(1)}^{0}u_{(2)}^{0}-u_{(1)}^{1}u_{(2)}^{1}-u_{(1)}^{3}u_{(2)}^{3}\nn\\
&=&\frac{\tilde{K}_{(1)}\tilde{K}_{(2)}-\sqrt{\tilde{K}_{(1)}^{2}-\Delta_{r} (\bar{K}_{(1)}^{2}+{K}_{(1)}^{2})}\sqrt{\tilde{K}_{(2)}^{2}-\Delta_{r} (\bar{K}_{(2)}^{2}+{K}_{(2)}^{2})}}{\Delta_{r} K_{(1)}K_{(2)}}\nn\\
&&-\frac{\bar{K}_{(1)}\bar{K}_{(2)}}{K_{(1)}K_{(2)}}, \label{4.2}
\ea
where
\ba
K_{(i)}&=&K|_{s=s_{i},j=j_{i}},\nn\\
\tilde{K}_{(i)}&=&\tilde{K}|_{s=s_{i},j=j_{i}},\nn\\
\bar{K}_{(i)}&=&\bar{K}|_{s=s_{i},j=j_{i}},,\hspace{2cm}i=1,2.
\ea
are the quantities corresponding to the particle 1 or 2.
One can easily see that for the case $\Lambda=0$ our result of the center-of-mass energy is the same as Kerr spacetime with spinning particles\cite{084025}. For $\Lambda=0$ and $a=0$, our result reduces to the situation of the Schwarzschild black hole with spinning particles\cite{105014}. And of course the case of the Kerr
black hole with spinless particles is recovered when we set $\Lambda=0$ and $s=0$\cite{111102}.

In the first sight, one may naively thought that $E_{cm}$ could diverge when the particles approach to the horizon since the value of $\Delta _r$ is zero at the horizon. However, the denominator of $E_{cm}$ could also be divergent. Therefore, we need to carefully analyze the asymptotic behavior of $E_{cm}$  when it approaches to horizon.  To this aim, we define
\ba
E_{0}=\tilde{K}_{(1)}\tilde{K}_{(2)}-\sqrt{\tilde{K}_{(1)}^{2}-\Delta_{r} (\bar{K}_{(1)}^{2}+{K}_{(1)}^{2})}\sqrt{\tilde{K}_{(2)}^{2}-\Delta_{r} (\bar{K}_{(2)}^{2}+{K}_{(2)}^{2})}, \label{4.4}
\ea
The first fraction of Eq. (\ref{4.2}) may be divergence when particles collide at the horizon $r=r_+$. In order to ensure if Eq. (\ref{4.2}) could be infinite at the horizon, we expand $E_{0}$ near the
horizon $r=r_+$
\ba
E_{0}=a+b(r-r_+)+...
\ea
where the first coefficients of the Taylor expansion $a$ reads
\ba
a=E_{0}|_{r=r_+}=\left[\tilde{K}_{(1)}\tilde{K}_{(2)}-\sqrt{\tilde{K}_{(1)}^{2}}\sqrt{\tilde{K}_{(2)}^{2}}\right]_{r=r_+}=0
\ea
On the other hand, from Eq. (\ref{4.4}), we can write that
\ba
b=\frac{dE_{0}}{dr}\Big|_{r=r_{+}}=\frac{K_{(2)}^{2}(\bar{K}_{(1)}^{2}+K_{(1)}^{2})+K_{(1)}^{2}(\bar{K}_{(2)}^{2}+K_{(2)}^{2})}{2\tilde{K}_{(1)} \tilde{K}_{(2)}}{\Delta }'_r
\ea
where ${\Delta }'_r=\frac{d{\Delta }_r}{dr}$. By using the expression of ${\Delta}_r$, and note that near the horizon $\Delta _{r}\sim r-r_+$(non-extremal) or $\Delta _{r}\sim (r-r_+)^2$(extremal). The Eq. (\ref{4.2}) at horizon $(r=r_+)$ can be written as
\ba
-g_{ab}u_{(1)}^{a}u_{(2)}^{b}=\frac{{\Delta'_r }  [K_{(2)}^{2}(\bar{K}_{(1)}^{2}+K_{(1)}^{2})+K_{(1)}^{2}(\bar{K}_{(2)}^{2}+K_{(2)}^{2})]}{2\tilde{K}_{(1)} \tilde{K}_{(2)}K_{(1)}K_{(2)}}-\frac{\bar{K}_{(1)}\bar{K}_{(2)}}{K_{(1)}K_{(2)}} \label{4.8}
\ea
Therefore, if we want the $E_{cm}$ blow up for the non-extremal black holes£¬ the only possibility is that $\tilde{K}_{(i)}=0$ or $K_{(i)}=0$, where $i=1,2$ at horizon. On the other hand we note that for the non-extremal Kerr-dS black hole, ${\Delta }'_r<0$ \cite{225006}; while, for the non-extremal Kerr-AdS black hole, the ${\Delta }'_r>0$. We will discuss these two cases separately in the following sections.

\section{Kerr-de Sitter black hole}\label{Kerr-dS}
In this section, we consider the spin particles are accelerated in Kerr-dS background($\Lambda>0$). Let $j=l+s$ as the total angular momentum, with $l$ being the orbital angular momentum, where $(l=j-s)$ \cite{105014}. Since $v^{a}$ is a timelike vector, we know that $\frac{dt}{d\tau}>0$ near the horizon $r=r_+$. Using Eq. (\ref{2.7}) and the normalization condition Eq. (\ref{3.23}), the relation between $v^{a}$ and $u^{a}$ can be solved
\ba
v^{(0)}&=&X^{-1}(1-\frac{Ms^{2}}{r^{3}})u^{(0)},\nn\\
v^{(1)}&=&X^{-1}(1-\frac{Ms^{2}}{r^{3}})u^{(1)},\nn\\
v^{(3)}&=&X^{-1}(1+\frac{2Ms^{2}}{r^{3}})u^{(3)}. \label{3.25}
\ea
where
\be
X=1-\frac{Ms^{2}}{r^{3}}[1+3(u^{(3)})^2]. \label{3.26}
\ee
The general form of 4-velocity $v^{a}$ of a spinning particle can be expressed as \cite{084025}
\be
v^{a}=\frac{dt}{d \tau }(\frac{\partial  }{\partial t})^{a}+\frac{dr}{d \tau }(\frac{\partial  }{\partial r})^{a}+\frac{d\varphi }{d \tau }(\frac{\partial  }{\partial \varphi})^{a}.\label{va}
\ee
Plugging Eqs. (\ref{3.6})- (\ref{3.9}) into Eq.(\ref{va}), we have
\ba
v^{(0)}&=&\sqrt{\frac{\Delta_{r} }{\Sigma }}(\frac{dt}{d \tau }-\frac{asin^2\theta }{\Xi } \frac{d\varphi }{d \tau }), \nn\\
v^{(1)}&=&\sqrt{\frac{\Sigma }{\Delta _r}} \frac{dr}{d \tau },\nn\\
v^{(3)}&=&\sqrt{\frac{\Delta _\theta}{\Sigma}}sin\theta(-a\frac{dt}{d \tau }+\frac{r^2+a^2}{\Xi }\frac{d\varphi }{d \tau }). \label{3.28}
\ea
Combining Eqs. (\ref{3.25}) and (\ref{3.28}), we can get
\ba
\frac{dt}{d\tau}&=&\frac{(a^2+r^2)(1-\frac{Ms^2}{r^3})\sqrt{\Delta _{r}}u^{(0)}+a(1+\frac{2Ms^2}{r^3})\Delta _{r}u^{(3)}}{X r\Delta _{r}}, \label{3.29}\\
\frac{dr}{d\tau}&=&\frac{\sqrt{\Delta _r}(1-\frac{Ms^2}{r^3})u^{(1)}}{X r}, \label{3.30}\\
\frac{d \varphi}{d\tau}&=&\frac{\Xi (a \sqrt{\Delta _{r}}(1-\frac{Ms^2}{r^3})u^{(0)}+ \Delta _{r}(1+\frac{2Ms^2}{r^3})u^{(3)})}{X r \Delta _{r}}. \label{3.31}
\ea
Substituting Eqs. (\ref{3.18}), (\ref{3.19}) and (\ref{3.26}) to Eq.(\ref{3.29}),we can write it as
\ba
\frac{dt}{d\tau}=\frac{(a^2+r^2)(r^3-Ms^2)\tilde{K}+a(r^3+2Ms^2)\Delta _{r}\bar{K}}{[r^3-Ms^2(1+(\frac{\bar{K}}{K})^2)]K r\Delta _{r}}
\ea
Since $\Delta _{r}|_{r=r_+}=0$, the numerator can be simplified as $(a^2+r^2)(r^3-Ms^2)\tilde{K}$ at the horizon, and note that $s\ll M$ \cite{084025,406}. The condition $dt/d\tau>0$ is equivalent to
\ba
\tilde{K}> 0.
\ea
Solving this equation gives us a upper limit of the total angular momentum $j$
\ba
j<\frac{3(3 M a s+\Lambda  a^4 r_{+}+\Lambda  a^2 r^3_{+}+3 a^2 r_{+}-\Lambda  a r^3_{+} s+3 a r_{+} s+3 r^3_{+})}{(\Lambda  a^2+3) (3 M s+3 a r-\Lambda s r^3_{+} )}=j_c
\ea
When $\Lambda=0$, this upper limit of total angular momentum $j_c$ is coincide with the critical angular momentum in Kerr black hole background\cite{084025}. Although in the first glance, the limit case $j=j_c$(which corresponds $\tilde{K}=0$) can not be obtained. However, similar in\cite{084025}, in the case of $j=j_c$, we still have $v^av_a<0$, namely the collision particles with critical angular momentum $j=j_c$ is still timelike.

\begin{figure}[!htb]
\subfigure[ͼ]{
\label{fig:subfig:a}
\includegraphics [width=0.4\textwidth]{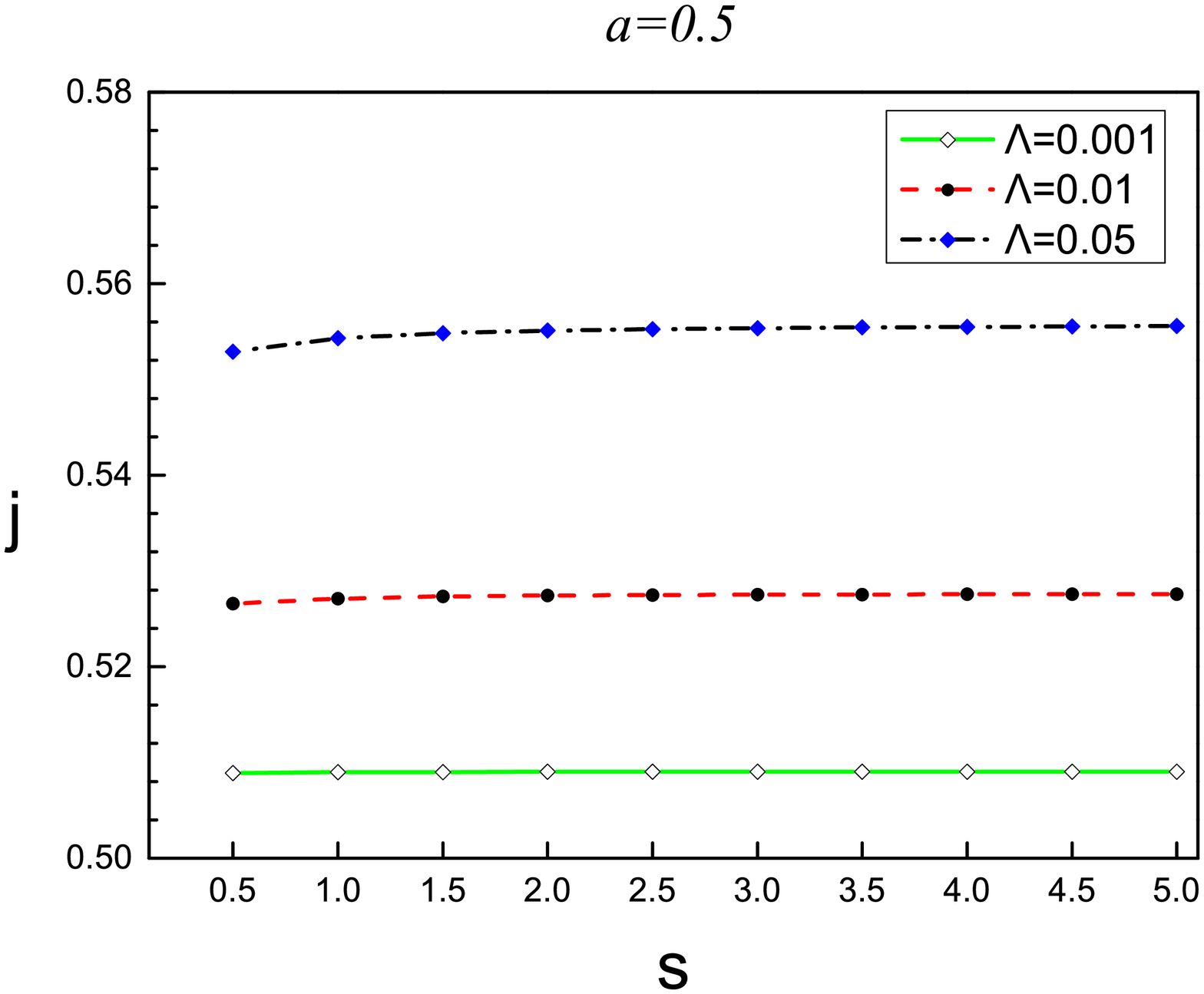}}
\subfigure[ͼ]{
\label{fig:subfig:b}
\includegraphics [width=0.4\textwidth]{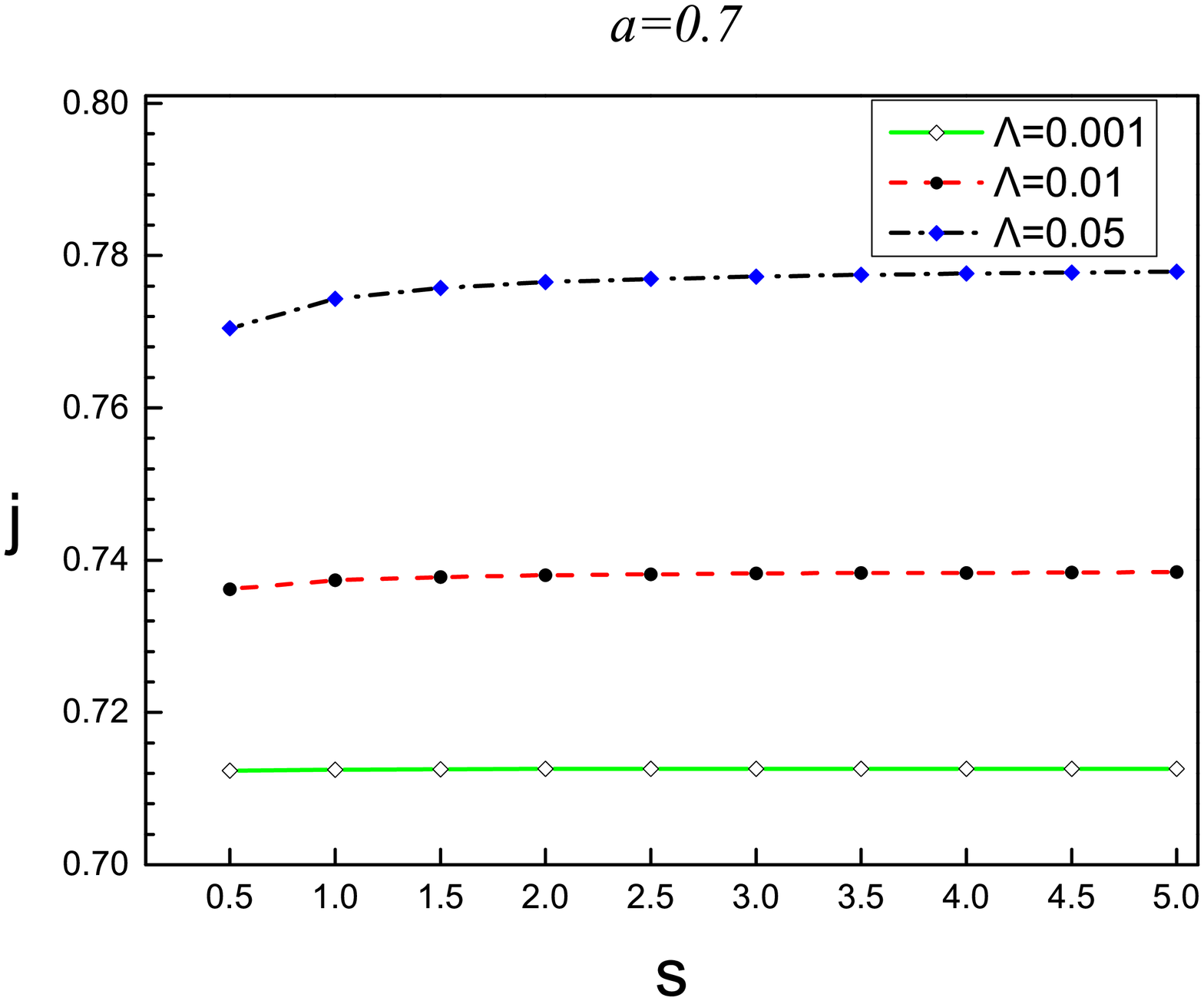}}
\caption{ \label{1} (a) Critical angular momentum $j$ as a function of spin $s$ for the Kerr-dS background with different value of $\Lambda$, ($a=0.5$). (b) Critical angular momentum $j$ as a function of spin $s$ for the non-extremal Kerr-dS background with different value of $\Lambda$, ($a=0.7$)}
\label{Fig.1}
\end{figure}

Fig.\ref{Fig.1} shows the effect of the different $a$ and cosmological constants $\Lambda$ on the critical angular momentum $j_c$. We can see that the $j_c$ becomes bigger as cosmological
constants $\Lambda$(spin of the black hole $a$)  increase.

Therefore, for a non-extremal Kerr-dS background, the collision particles with critical angular momentum $j=j_c$ has $\tilde{K}=0$. By using Eq.(\ref{4.8}), we have a blow up $-g_{ab}u_{(1)}^{a}u_{(2)}^{b}$, which in turn implies an arbitrarily high center-of-mass energy $E_{cm}$.

Now, we verify that for the critical situation whether the spinning particle can really reach the horizon. It is found that the radial turning points occurred at the point of $u^{r}=0$\cite{105014}. Since $u^{r}=\frac{P^{r}}{m}$, at the turning points, $(P^{r})^2=0$. Using Eqs.(\ref{3.18}), (\ref{3.19}) and (\ref{3.24}), the non-vanishing components of the momentum are
\ba
\frac{P^{t}}{m}&=&\frac{(r^2+a^2)\tilde{K}+a\bar{K}\Delta _{r}}{\Delta_{r}Kr}, \\
\frac{P^{\phi }}{m}&=&\frac{(a\tilde{K}+\bar{K}\Delta _{r})\Xi }{\Delta_{r}Kr},  \\
\left(\frac{P^{r}}{m}\right)^{2}&=&\frac{\tilde{K}^{2}-\Delta _{r}(\bar{K}^{2}+K^{2}) }{K^{2}r^{2}}. \label{5.17}
\ea

Note that for the non-extremal Kerr-dS black hole, $\Delta _{r}\sim (r-r_+)$ and $\tilde{K}=0$, we get $(\frac{P^r}{m})^2\Big|_{r=r_{+}}=0$. Obviously, the spinning particles can reach the horizon and the collision energy $E_{cm}$ is divergent near the horizon. Therefore, the collision center-mass-energy $E_{cm}$ blow up at $r=r_+$ when the critical collision angular momentum $j=j_c$ satisfied even for the non-extremal Kerr-dS black hole.

On the other hand, from Eq.(\ref{4.8}), it seems that the center-mass-energy $E_{cm}$ could also being divergent when $K=0$. However, in this case, the particle can not approach to the horizon because of $(\frac{P^r}{m})^2\Big|_{r=r_{+}}\neq0$ when $K=0$.

\section{Kerr-anti-de Sitter black hole}\label{Kerr-AdS}
Now we turn to Kerr-anti-de Sitter background. In this case which is unlike Kerr-dS background, the $\Delta'_r\geq0$. The $E_{cm}$ blow up means that $\tilde{K}_{(i)}=0$ or $K_{(i)}=0$, where $i=1,2$. This condition can be realized when $r=r_+$. However, in order to insure the particle can escape to infinite, we need to require that the derivative of $(P^r)^2$ respect to $r$ must be positive at the horizon $r_+$ \cite{225006}.
\ba
\frac{d(\frac{P^r}{m})^2}{dr}\Big|_{r=r_+}>0 \label{dpr}
\ea
Substituting Eq.(\ref{5.17}) to the above formula, we can get
\ba
\frac{d(\frac{P^r}{m})^2}{dr}\Big|_{r=r_+}=\frac{2\tilde{K}\tilde{K}'-\Delta'_r(\bar{K}^2+K^2) }{2Kr(K'r+K)}
\ea
Since for the non-extremal Kerr-anti-de Sitter black hole, the $\Delta'_r>0$, then the $\frac{d(\frac{P^r}{m})^2}{dr}\Big|_{r=r_+}$ is always negative at the horizon, which means the center of the energy $E_{cm}$ can not reach arbitrary high. For the extremal Kerr-AdS black hole $\Delta'_r=0$, which means $\frac{d(\frac{P^r}{m})^2}{dr}\Big|_{r=r_+}=0$, however, we found that $\frac{d^2(\frac{P^r}{m})^2}{dr^2}>0$, the collision particles can still escape to infinite. This phenomenon also found for the extremal Kerr-AdS black hole with spinless collision particles\cite{225006}. From these analysis, we get a conclusion that $E_{cm}$ can only blow up for extremal Kerr-AdS black hole.

\section{Conclusions}\label{conclusion}
In this paper, we have analyzed the possibility that Kerr-dS and Kerr-AdS black holes could act
as accelerators for spinning particle. We find that the result is very different from the case of Kerr black holes
due to the existence of the non-vanishing cosmological constant. It turns out that collision of two particles in the outer horizon with the critical spinning angular momentum $j=j_c$ can reach arbitrary high center-of-mass energy. Besides, for the case of the Kerr black hole, it has to be extremal. However, for the case of the Kerr-dS black hole, it does not need to be extremal. Hence, non-extremal Kerr-dS black holes
could also serve as particle accelerators with arbitrarily high center-of-mass energy $E_{cm}$, which is very different
from the cases of the Kerr and Kerr-AdS black holes. By detailed analysis, the sign of $\Delta'_r$ is different for Kerr-AdS and Kerr-dS case, and this difference makes exactly why the Kerr-dS case is so distinct with other backgrounds.

\begin{acknowledgements}
This work is supported by NSFC with No.11775082  and
the Fundamental Research Funds for the Central University of China. The authors would like thank Prof. Sijie Gao for helpful discussions.

\end{acknowledgements}

\end{document}